\documentstyle[tighten,aps,eqsecnum]{revtex}

\newcommand{\dif}{{\rm d}}
\newcommand{\Dif}{{\rm D}}
\newcommand{\Man} {{\cal M}}
\newcommand{\bgamma}{\bbox{\gamma}}
\newcommand{\bxi}{\bbox{\xi}}
\newcommand{\bomega}{\bbox{\omega}}

\begin{document}

\draft

\title{Einstein-Cartan-Maxwell Theory with Scalar Field\\
	through a Five-Dimensional Unification}

\author{{\sc Christopher Kohler}\thanks{e-mail: {\sf c.kohler@rz.uni-sb.de}}}
	
\address{Fachrichtung Theoretische Physik, Universit\"at des Saarlandes,\\
	Postfach 151150, D-66041 Saarbr\"ucken, Germany}
	
\date{\today}
	
\maketitle

\begin{abstract}
	A modification of Kaluza-Klein theory is proposed in which, as a result
of a symmetry breaking, five-dimensional space-time is partially parallelized 
implying the appearance of torsion fields. A naturally chosen action 
functional leads to the Einstein-Cartan-Maxwell theory where the 
electromagnetic field strength is represented by the fifth component of
the torsion 2-form. Incorporation of a scalar field reveals that 
four-dimensional space-time torsion is not induced by the 
scalar field.	
\end{abstract}



\section{Introduction}

The traditional Kaluza-Klein models \cite{kalu,kle} are restricted to 
Riemannian geometry of five-dimensional (5-d) space-time in that the 
condition of vanishing torsion is imposed {\em a priori}. In view of the 
various generalizations of Einstein's general theory of relativity to 
non-Riemannian space-time geometries (see Ref.\ \cite{heh} for a review), 
it is a natural task to investigate higher dimensional unifications based 
on such alternative theories of gravitation. The incorporation of torsion 
degrees of freedom is the most important modification. Several 
authors have already studied Kaluza-Klein models in space-times with torsion 
using, e.g., 5-d Riemann-Cartan space \cite{kal,ger,ger2,zha}, 5-d teleparallel 
space \cite{sch,lee}, or special definitions of the space-time geometry 
\cite{kop,kat,kat2,bat}. In this article, we address ourselves to the problem 
of deriving pure Einstein-Cartan-Maxwell theory from a 5-d gravitation
within a closed formalism.

We arrive at this goal by means of the imposition of constraints on the 5-d 
space-time connection which are suggested by the local symmetries of the 
geometry: 5-d gravitation possesses local $SO(4,1)$ symmetry whereas the 
dimensionally reduced gravitation has only local $SO(3,1)$ symmetry of the 
external space-time. The dimensional reduction thus involves a symmetry 
breaking. The basic idea of this article is to break the symmetry 
{\em before} the dimensional reduction --- {\em on the kinematical level}. 

In section 2, we will show that this approach is consistent with two restrictions on 
the 5-d connection: ($i$) The normalized Killing vector field along the 
internal $S^1$-manifold is a {\em parallel} vector field with respect to 
the 5-d connection. ($ii$) The parallel transport around the $S^1$-manifold 
is integrable. 

Both restrictions concern the distant parallelism of space-time and 
necessarily imply the appearance of torsion fields. Since the parallel 
transport in 4-d space-time is not restricted, these geometries shall 
be referred to as {\em semi-teleparallel}.

In section 3 of this article, we construct an action functional for 
semi-teleparallel geometries. The choice of the action is a natural 
one in the sense that the construction leads in the extreme cases 
of pure Riemann-Cartan geometry and teleparallel geometry to the 
standard actions usually employed.

The subject of section 4 is the application of the chosen action functional 
to the semi-teleparallel Kaluza-Klein geometry without scalar field. 
After dimensional reduction, the Einstein-Cartan-Maxwell theory is 
exactly reproduced.

In section 5, we investigate the incorporation of a scalar field into 
the semi-teleparallel Kaluza-Klein model. In contrast to the usual 
approaches to scalar-tensor gravity with torsion, we do not obtain the 
result that space-time torsion is induced by the scalar field.

The final section contains a discussion of the approach presented in this 
article.


\section{Geometrical framework}

In this section, we establish the basic properties of a semi-teleparallel 
geometry. We assume the 5-d space-time manifold in Kaluza-Klein theory to 
have a direct product topology $\Man\times S^1$ where $\Man$ is the external 
4-d space-time manifold and the circle $S^1$ is the internal manifold. We 
further assume that on $\Man\times S^1$ a pseudo-Riemannian metric $\bgamma$ 
of signature ($-++++$) is defined which admits a spacelike Killing vector 
field $\bxi$ the integral curves of which are the $S^1$-manifolds.

Kaluza-Klein theory starts with general relativity on $\Man\times S^1$. The 
corresponding action functional possesses local $SO(4,1)$ invariance. The 
dimensionally reduced action, however, has only local $SO(3,1)$ invariance 
on $\Man$, besides the electromagnetic $U(1)$ invariance. In this article, 
we modify Kaluza-Klein theory in that we formulate a theory of gravitation 
on $\Man\times S^1$ that possesses local $SO(3,1)$ invariance defined on 
$\Man$ from the outset. The corresponding space-time geometry can be 
obtained through a symmetry breaking as described in the following.

We first consider a linear connection on $\Man\times S^1$. It is given by 
a $gl(5,{\bf R})$-valued connection 1-form \cite{kob}
\begin{equation}\label{2.1}
	\bomega = \left( e^A_M \partial_P e^M_B 
	+ e^A_M e^N_B \Gamma^M_{NP} \right) \dif x^P E^B_A .
\end{equation}
Here, $e^M_A$ are the components of the frame $e_A = e^M_A\partial_M$ 
where $A,B,\ldots = 0,1,2,3,5$ are frame indices and $M,N,\ldots = 0,1,2,3,5$ 
are indices of a local coordinate system $x^M$. (Explicit indices in 
parentheses will be frame indices in the following.) $e^A_M$ is the inverse 
matrix of $e^M_A$ forming the cobasis $e^A = e^A_M\dif x^M$. The functions 
$\Gamma^M_{NP}$ in Eq.\ (\ref{2.1}) are the Christoffel symbols and 
$E^A_B$ are the generators of $GL(5,{\bf R})$ satisfying the Lie algebra
\[
	[ E^A_B, E^C_D ] = \delta^A_D E^C_B - \delta^C_B E^A_D.
\]

The metric $\bgamma$ on $\Man\times S^1$ defines an equivalence class of 
orthonormal frames where $\gamma_{MN} = e^A_M e^B_N \eta_{AB}$ with the 
5-d Minkowski metric $\eta_{AB} = \mbox{diag}(-1,1,1,1,1)$. If the 
connection 1-form (\ref{2.1}) is restricted to such frames, we can perform 
the following decomposition:
\begin{equation}\label{2.2}
	\bomega = -\frac{1}{2} e^M_A e^N_B \Dif_P \gamma_{MN} \dif x^P E^{(AB)} +
	\left( e_{AM} \partial_P e^M_B + e_{AM} e^N_B \Gamma^M_{NP} \right) 
	\dif x^P E^{[AB]},
\end{equation}
where $\Dif_M$ represents the covariant derivative with respect to 
$\Gamma^M_{NP}$ and frame indices are raised and lowered using $\eta_{AB}$. 
The generators $E^{(AB)}$ and $E^{[AB]}$ are the symmetric and antisymmetric 
part of $E^{AB} \equiv \eta^{AC} E^B_C$, respectively. $\Dif_P\gamma_{MN}$ is 
the well known tensor of nonmetricity the vanishing of which is equivalent 
to the reduction of $\bomega$ to the Lorentz connection
\begin{equation}\label{2.3}
	\bomega = \frac{1}{2}\omega_{AB} J^{AB} = \frac{1}{2}\left( e_{AM} 
	\partial_P e^M_B + e_{AM} e^N_B 
	\Gamma{}^M_{NP} 
	\right)\dif x^P  J^{AB},
\end{equation}
where $J^{AB} \equiv 2E^{[AB]}$ are the generators of $SO(4,1)$. The geometric
interpretation of the condition $\Dif_P\gamma_{MN} = 0$ is that the parallel 
transport of $\bgamma$ is integrable, that is, $\bgamma$ is a parallel tensor 
field. 

In Kaluza-Klein theory, it is not sufficient to have a metric on $\Man\times 
S^1$; the existence of the Killing vector $\bxi$ gives a further 
constraint on the geometry. We shall treat the Killing vector on an equal 
footing with the metric $\bgamma$ in the sense that we use the Killing 
vector to reduce the $SO(4,1)$ connection (\ref{2.3}) to an $SO(3,1)$ 
connection in the same way as the metric $\bgamma$ was used to reduce the
$GL(5,{\bf R})$ connection (\ref{2.1}) to the $SO(4,1)$ connection 
(\ref{2.3}). For that, we note that the Killing vector $\bxi$, together 
with the metric $\bgamma$, defines a restricted equivalence class of 
orthonormal frames by requiring that $e_{(5)}$ lies in the direction 
of $\bxi$, that is, $e_{(5)}$ is the normalized Killing vector. Using these
frames, Eq.\ (\ref{2.3}) can be decomposed as
\begin{equation}\label{2.4}
	\bomega =  e_{aM} \Dif_P e^M_{(5)} \dif x^P J^{a5} + \frac{1}{2}
	\left( e_{aM} \partial_P e^M_b + 			
	e_{aM} e^N_b \Gamma{}^M_{NP} \right) \dif x^P J^{ab} ,
\end{equation}
where $a,b,\ldots = 0,1,2,3$. If the connection is such that $\omega_{a(5)P} =
e_{aM}\Dif_P e^M_{(5)}$ vanishes, which is equivalent to
\begin{equation}\label{2.5}
	\Dif_P e^M_{(5)} = 0 ,
\end{equation}
then the connection reduces to the $SO(3,1)$ connection 
\begin{equation}\label{2.6}
	\bomega = \frac{1}{2}\left( e_{aM} \partial_P e^M_b + e_{aM} e^N_b 
	\Gamma{}^M_{NP} \right) \dif x^P J^{ab} .
\end{equation}
The condition (\ref{2.5}) means that $e_{(5)}$ is a parallel vector 
field, that is, the internal circles represent autoparallels.

While the vanishing of the nonmetricity can always be achieved by choosing 
the connection to be the Levi-Civita connection with respect to $\bgamma$, 
the condition (\ref{2.5}) can in general only be fullfilled if the 
connection possesses torsion. In this case, however, we must further 
restrict the geometry by requiring that the Killing vector be an infinitesimal 
affine transformation \cite{kob}. This is a natural requirement since it
implies that --- as in Kaluza-Klein theory --- the Christoffel symbols 
are independent of a coordinate on $S^1$, which shall be $\theta$ with 
range $[0,1]$ in the following. ($x^\mu$, $\mu = 0,1,2,3$, will be 
coordinates on $\Man$.) The Killing vector field is then given by $\bxi =
\partial_\theta$.

Although Eq.\ (\ref{2.6}) represents an $SO(3,1)$ connection, the symmetry 
breaking is not yet completed since we still have the freedom to perform 
local $\theta$-dependent $SO(3,1)$ transformations of the frames. In order 
to remove this symmetry, we single out frames $e_a$ that are parallel 
along $S^1$, that is, the connection is such that $\omega_{ab5} = 0$ in
these frames. The fact that the Christoffel symbols do not depend on $\theta$ 
and the frames have to be single-valued on $S^1$ imposes strong restrictions 
on such frames. The simplest choice are $\theta$-independent frames. 
These are determined up to local $SO(3,1)$ transformations on $\Man$, 
which is the required symmetry. 

To summarize, we have shown that the breaking of the local $SO(4,1)$ symmetry 
on $\Man\times S^1$ down to a local $SO(3,1)$ symmetry that is defined on 
$\Man$ specifies a connection for which the components $\omega^a{}_{(5)M}$ and 
$\omega^a{}_{b5}$ of the connection 1-form vanish in a basis where $e_{(5)}$ 
is the normalized Killing vector field and the basis vectors $e_a$ do not 
depend on $\theta$. Hence, in this frame only the components 
$\omega^a{}_{b\mu}$ are nonvanishing. This means that, first, the parallel 
transport of a vector pointing in the fifth dimension is integrable and, 
secondly, the parallel transport of an arbitrary vector along the fifth 
dimension is integrable. Since the parallel transport does not depend on 
$\theta$, this implies that the components $R^{(5)a}{}_{MN}$ and 
$R^{ab}{}_{5\mu}$ of the curvature 2-form, defined by $R^{AB} = 
\dif \omega^{AB} + \omega^A{}_C\wedge \omega^{CB}$, vanish. This geometry 
is similar to the teleparallel (or Weitzenb\"ock) geometry. However, the 
parallelization is performed here only partially justifying the term 
semi-teleparallel geometry. The corresponding connection also appears, as a 
special case, in Ref.\ \cite{kop}.


\section{Action functional}

The purpose of this section is to construct an appropriate action functional 
describing the dynamics of the semi-teleparallel geometry. This action should 
be as close as possible to the usual gravitational actions. Since the 
semi-teleparallel geometry lies ``in between'' the pure Riemann-Cartan and the 
teleparallel geometry, we seek an action which ``interpolates'' those of the 
Einstein-Cartan and the teleparallel gravitation.

Our starting point is the action for the Einstein-Cartan gravitation in five 
dimensions,
\begin{equation}\label{3.1}
	\tilde{S} [\tilde{e}{}^A_M,\tilde{\omega}{}^A{}_{BM} ] =
	  \int_{\Man\times S^1} \dif^5 x \sqrt{-\gamma} \tilde{R} ,
\end{equation}
where $\gamma$ is the determinant of the metric tensor $\gamma_{MN}$ and 
$\tilde{R} = \tilde{R}{}^{AB}{}_{AB}$ is the 5-d scalar curvature. 
The tildes in (\ref{3.1}) indicate that the respective quantities are not 
yet those of the semi-teleparallel geometry. 
A possible procedure for obtaining an action for the teleparallel gravitation
consists in the replacement of the components $\tilde{\omega}^A{}_{BM}$ in
(\ref{3.1}) by the Levi-Civita connection components 
$\overcirc{\omega}^A{}_{BM}$ and in the subsequent interpretation of the 
frame $\tilde{e}{}^A_M$ as a teleparallel frame $e^A_M$, that is, the 
connection components $\omega^A{}_{BM}$ are zero. The resulting action is 
equivalent to the Einstein action, the underlying geometry, however, is 
different from the one of general relativity.

The teleparallel geometry corresponds to a reduction of a Lorentz connection
to a $\{ 1\}$ connection. The fact that the semi-teleparallel geometry 
corresponds to a reduction of an $SO(4,1)$ connection to an $SO(3,1)$ 
connection suggests that in order to construct an action one should replace 
only the components $\tilde{\omega}{}^a_{(5)M}$ by the Levi-Civita connection 
components $\overcirc{\omega}{}^a{}_{(5)M}$. The geometry must then be 
interpreted as a semi-teleparallel geometry, that is, $\tilde{e}{}^A_M$ is 
a semi-teleparallel frame $e^A_M$ with $\omega^a{}_{(5)M} = 0 = 
\omega^a{}_{b5}$. Using the general relation 
\begin{equation}\label{3.1a}
	R = \overcirc{R} + \frac{1}{4} T^{ABC}T_{ABC} + \frac{1}{2}
		T^{ABC}T_{CBA} + T^B{}_{BA}T^{CA}{}_C - 2\overcirc{D}_A T^B{}_B{}^A
\end{equation}
between the full scalar curvature $R$, the Riemannian scalar curvature 
$\overcirc{R}$, and the torsion tensor $T^A{}_{BC}$ defined by $T^A = 
\dif e^A + \omega^A{}_B \wedge e^B$, we obtain, as a result of these 
substitutions, an action that no longer depends on $\omega^a{}_{b5}$ and 
$\omega^a{}_{(5)M}$: 
\begin{equation}\label{3.2}
	S[e^A_M,\omega^a{}_{b\mu}] = \int_{\Man\times S^1} \dif^5x \sqrt{-\gamma}
		\left( \overcirc{R} + \frac{1}{4} T^{abc}T_{abc}
		+ \frac{1}{2}T^{abc}T_{cba} + T^b{}_{ba}T^{ca}{}_c \right),
\end{equation}		 			
where a surface term in the integrand has been dropped.

The functional (\ref{3.2}) is the action we will use for the semi-teleparallel
formulation of the Kaluza-Klein model. The action can be written in a more 
suitable form if we eliminate  $\overcirc{R}$ in (\ref{3.2}) with the help of 
Eq.\ (\ref{3.1a}) yielding
\begin{equation}\label{3.4}
	S[e^A_M,\omega^a{}_{b\mu}] = \int_{\Man\times S^1} \dif^5x \sqrt{-\gamma}
		\left( {}^{(4)}R - \frac{1}{4} T^{(5)ab}T_{(5)ab} 
		- 2 T^{(5)}{}_{(5)a}T^{ba}{}_b \right),
\end{equation}
where we have replaced $R$ by the 4-d scalar curvature ${}^{(4)}R = 
R^{ab}{}_{ab}$ which is 
possible because the components $R^{a(5)}$ of the curvature 2-form vanish. 
Moreover, we have discarded the divergence appearing in Eq.\ (\ref{3.1a}) 
which leads to a surface term in (\ref{3.4}). 


\section{Einstein-Cartan-Maxwell theory}

In order to derive from the action (\ref{3.4}) an effective 4-d action through
dimensional reduction, we have to specify the basis $e_A$. A natural choice 
for $e_{(5)}$ is $ e_{(5)} = \partial_\theta$. As for $e_a$, we make the 
ansatz $ e_a = e^\mu_a \left(\partial_\mu -  A_\mu\partial_\theta\right)$ 
where $e^{\mu}_a$ and $A_\mu$ are functions on $\Man$. The corresponding 
cobasis is 
\begin{equation}\label{4.1}
	e^a = e^a_\mu \dif x^\mu , \quad e^{(5)} = \dif\theta + A_\mu \dif x^\mu.
\end{equation}
The parameterization of this cobasis leads to the metric usually employed 
in the Kaluza-Klein model,
\begin{equation}\label{4.2}
	\gamma_{MN} =  \left(
		\begin{array}{cc}
		g_{\mu\nu} +  A_{\mu}A_{\nu} &  A_{\mu} \\
		 A_{\nu}  &  1
		\end{array}	
		\right) ,
\end{equation}
where $g_{\mu\nu} = e^a_\mu e^b_\nu \eta_{ab}$.
It should be remarked that in the present case $e_{(5)}$ is fixed and 
the $e_a$ are only determined up to local 4-d Lorentz transformations that 
are independent of $\theta$. Inserting the cobasis (\ref{4.1}) into the 
action (\ref{3.4}) we obtain
\begin{equation}\label{4.3}
	S = \int_{\Man\times S^1} \dif^5x \sqrt{-\gamma} \left( {}^{(4)}R - 
	\frac{1}{4} F_{\mu\nu} F_{\rho\sigma} g^{\mu\rho} g^{\nu\sigma} \right),
\end{equation}
where $F_{\mu\nu} \equiv 2\partial_{[\mu} A_{\nu]}$. Since $\sqrt{-\gamma} = 
\sqrt{-g}$ and the integrand does not depend on $\theta$, we can integrate 
over the internal $S^1$ yielding
\begin{equation}\label{4.4}
	S =  \int_{\Man} \dif^4x \sqrt{-g} \left( {}^{(4)}R - \frac{1}{4} 
	F_{\mu\nu} F^{\mu\nu}
	 \right),
\end{equation}
where $F^{\mu\nu} \equiv F_{\rho\sigma} g^{\mu\rho} g^{\nu\sigma}$.
The functional (\ref{4.4}) is the action of the Einstein-Cartan-Maxwell 
theory. It should be emphasized that here the field strength is represented 
by the fifth component of the torsion tensor, $F_{\mu\nu} = 
T^{(5)}{}_{\mu\nu}$. Another argument in favor of this interpretation 
of the field strength was given in Ref.\ \cite{koh} in connection with an
analysis of the formulation of the Aharonov-Casher effect.			
		

\section{Incorporation of a scalar field}

The cobasis (\ref{4.1}) is not the most general parameterization of the 
Kaluza-Klein geometry. We can introduce a scalar field $\phi (x^\mu)$ 
which leads to a scale factor of the internal $S^1$-manifold. Choosing 
the parameterization $e_{(5)} = e^{-\phi}\partial_\theta$, the cobasis 
is given by
\begin{equation}\label{6.1}
	e^a = e^a_\mu \dif x^\mu, \quad e^{(5)} = e^{\phi}\left( \dif\theta +
	  A_\mu 	\dif x^\mu \right) .
\end{equation}
Inserting (\ref{6.1}) into the action (\ref{3.4}), we obtain after dimensional 
reduction
\begin{equation}\label{6.2}
	S = \int_{\Man} \dif^4x 	\sqrt{-g} \left( e^\phi \;{}^{(4)}R - 
	\frac{1}{4} e^{3\phi} F_{\mu\nu} F^{\mu\nu}+ 2e^\phi\, \partial_\mu 
	\phi\, T^\rho{}_{\nu\rho}\, g^{\mu\nu} \right) .
\end{equation}
There is an apparent coupling of the scalar field and the torsion tensor 
described by the last term in the action. However, the scalar curvature 
${}^{(4)}R$ contains a divergence $2g^{\mu\nu} \overcirc{D}_\mu 
T^\rho{}_{\nu\rho}$ which, after a partial integration, cancels the last 
term exactly. Thus, we obtain
\begin{equation}\label{6.3}
	S = \int_{\Man} \dif^4x 	\sqrt{-g} \left\{ e^\phi \left[ 
	{}^{(4)}\overcirc{R} + g^{\mu\nu} \left(
	K^\rho{}_{\mu\nu} K^\sigma{}_{\rho\sigma} - K^\rho{}_{\mu\sigma}
	 K^\sigma{}_{\rho\nu} \right) \right] - \frac{e^{3\phi}}{4} F_{\mu\nu} 
	 F^{\mu\nu} \right\} ,
\end{equation}
where 
\[
	K^\mu{}_{\nu\rho} \equiv \frac{1}{2} \left[ T^\mu{}_{\nu\rho} - 
	g^{\mu\sigma}
	\left( g_{\rho\lambda} T^\lambda{}_{\sigma\nu} + g_{\nu\lambda}
	 T^\lambda{}_{\sigma\rho} \right)\right]
\]
is the 4-d contortion tensor and ${}^{(4)}\overcirc{R}$ is the scalar 
curvature of the 4-d metric $g_{\mu\nu}$. The action (\ref{6.3}) shows 
that torsion is not induced by the scalar field; variation with respect 
to the torsion tensor leads to field equations which imply a vanishing 
torsion in the absence of spinning matter. This stands in contrast to 
some other approaches to gravitation with torsion and Brans-Dicke field 
\cite{ger3,kim,kim2} where the torsion acquires a contribution from the 
Brans-Dicke scalar even in the absence of matter.

The factor $e^{\phi}$ in the gravitational part of the action (\ref{6.3}) 
can be removed by introducing a different parameterization of the cobasis. An 
appropriate form is
\begin{equation}\label{6.4}
	\hat{e}{}^a =e^{-\frac{1}{3}\phi} e^a_\mu \dif x^\mu , \quad
	\hat{e}{}^{(5)} = e^{\frac{2}{3}\phi}\left(\dif\theta + A_\mu 
	\dif x^\mu\right)
\end{equation}
corresponding to a Weyl-factor $e^{-\frac{2}{3}\phi}$ in the metric. In this 
case, the action (\ref{3.4}) yields
\begin{equation}\label{6.5}
	S = \int_{\Man} \dif^4x 	\sqrt{-g} \left( {}^{(4)}R - \frac{1}{4} 
	e^{2\phi} F_{\mu\nu} F^{\mu\nu} -\frac{2}{3} \partial_\mu\phi\, 
	\partial_\nu\phi\, g^{\mu\nu} \right) .
\end{equation}
			

\section{Discussion}

In this article, we have put forward a modification of the Kaluza-Klein 
model which unifies 4-d Einstein-Cartan gravitation with electromagnetism 
and a scalar field. The main ingredient was the introduction of a 
geometry, which was called semi-teleparallel geometry, relying on a connection 
that parallelizes the direction of the space-time manifold singled out by a 
vector field. In a certain sense, we did not introduce new structures since 
the constraints on the space-time geometry were traced back to the dimensional 
reduction procedure and to the Killing vector field which already exist 
in the original Kaluza-Klein unification.

Besides this unification scheme with torsion, there are at least two 
alternative approaches. 

The first approach, which is treated in the literature \cite{kal,ger2,zha}, 
consists in the use of 5-d Einstein-Cartan gravitation as starting point. 
This procedure also leads to the Einstein-Cartan-Maxwell theory. However, 
there emerge extra fields --- the components of the torsion tensor involving 
the fifth dimension --- which have to be interpreted physically. These extra 
fields vanish in the vacuum by their field equations, but they will play a 
role when spinning matter is present. 

A possible second approach starts from 5-d Einstein-Cartan gravitation 
where the components of the torsion tensor involving the fifth dimension 
are set to zero {\em a priori}. This method leads to the same results
as obtained in this article. However, the geometry of space-time is 
completely different. Moreover, there is no obvious motivation to constrain 
a part of the torsion tensor. Generally, the imposition of the constraint 
of a vanishing torsion tensor is unnatural since it is not connected with 
symmetry properties --- as it is, e.g., in the case of vanishing 
nonmetricity. In the approach proposed in this article, the torsion is 
not restricted kinematically; it is the curvature that is constrained, 
and this follows from a symmetry breaking naturally implied by the 
dimensional reduction.

The concept of semi-teleparallel geometry may also be applied to the higher 
dimensional unification of gravitation with Yang-Mills theory. Furthermore, 
the concept could even be used in 4-d space-time. These generalizations
are presently under study.		
				

\end{document}